\documentclass{ws-procs9x6}

\newcommand{\beeq}{\begin{equation}}
\newcommand{\eneq}{\end{equation}}
\newcommand{\beeqa}{\begin{eqnarray}}
\newcommand{\eneqa}{\end{eqnarray}}

\begin{document}

\title{
{\vspace{-1.2em} \parbox{\hsize}{\hbox to \hsize
{\hss \normalsize UPRF-2003-24}}} \vglue 0.3 cm 
The Study of the Continuum Limit of the Supersymmetric Ward-Takahashi Identity for $N=1$ Super Yang-Mills Theory }
\author{A.~Feo}

\address{School of Mathematics, Trinity College, Dublin 2, Ireland and \\
Dipartimento di Fisica, Universit\`a di Parma and INFN Gruppo Collegato di Parma, 
Parco Area delle Scienze, 7/A, 43100 Parma, Italy }

\maketitle

\abstracts{The one-loop corrections to the supersymmetric Ward-Takahashi identity (WTi) are investigated in 
the off-shell regime in the Wilson formulation of the discretized $N=1$ Super Yang-Mills (SYM) theory.
The study of the continuum limit as well as the renormalization procedure for the supercurrent 
are presented.}

\section{Introduction}
Recently, there have been a number of interesting results in lattice supersymmetry, as for example, 
in the two dimensional Wess-Zumino model\cite{wesszumino}, four dimensional N=1 SYM theory\cite{feo}, 
with chiral fermions or not, as well as other supersymmetric theories\cite{others}.
Here, to formulate supersymmetry on the lattice we follow the ideas of Curci and Veneziano\cite{curci}. 
What they propose is to give up manifest supersymmetry on the lattice, and instead, 
to restore it in the continuum limit, by tunning the 
bare coupling $g_0$ and the gluino mass to the supersymmetric point which also coincides with the chiral point.
In the Wilson formulation of Curci and Veneziano lattice supersymmetry can also be investigated by means of the WTi.
On the lattice, the WTi contains explicit supersymmetric breaking terms and the 
supersymmetric limit is defined to be the point in the parameter space where these breaking terms
vanish and the continuum supersymmetric WTi is recovered.
Nice numerical results\cite{farchioni}, are already in the literature, but still the study of 
the continuum limit of the supersymmetric WTi was missing.
Here we want to shed some light on how to deal with this difficult issue, illustrating 
a recent result\cite{ward}, that shows that, it is possible to write down the renormalized supersymmetric 
WTi and the general procedure to determine the renormalization coefficient for the supercurrent, $Z_T$.

\section{Renormalized supersymmetric WTi on the lattice} 
The starting point of our calculation is the renormalized supersymmetric WTi on the lattice, 
which has been introduced in Ref.~\refcite{ward}, 
\beeqa
&& Z_S \big< O \nabla_\mu S_\mu(x) \big> + Z_T \big< O \nabla_\mu T_\mu(x) \big> - 
2 (m_0 - \tilde{m}) Z_{\chi}^{-1} \big< O \chi^R(x) \big> + \nonumber \\
&& Z_{CT} \big< \frac{\delta O} {\delta \bar \xi(x)}|_{\xi = 0} \big> - 
Z_{GF} \big< O \, \frac{\delta S_{GF}}{\delta \bar \xi(x)}|_{\xi = 0} \big> -
Z_{FP} \big< O \, \frac{\delta S_{FP}}{\delta \bar \xi(x)}|_{\xi = 0} \big> + \nonumber \\ 
&& \sum_j Z_{B_j} \big< O B_j \big>=0 \, ,
\label{renorm}
\eneqa
where $\xi(x)$ is a localized transformation parameter, 
$ S_\mu(x) = -\frac{2 i}{g_0} \, \mbox{Tr} \, \big( G_{\rho \tau}(x) \sigma_{\rho \tau} \gamma_\mu \lambda(x) \big)$,
is a local definition of the lattice supercurrent which mixes with
$ T_\mu(x) = -\frac{2}{g} \, \mbox{Tr} \, \big( G_{\mu \nu}(x) \gamma_\nu \lambda(x) \big) $.
$\chi(x) = \frac{i}{g_0} \, \mbox{Tr} \, \big( G_{\rho \tau}(x) \sigma_{\rho \tau} \lambda(x) \big) $
is the gluino mass term, 
$\nabla_\mu $ is the symmetric lattice derivative,
$G_{\rho \tau}(x)$ is the clover plaquette operator and $\sigma_{\rho \tau} = \frac{i}{2} [\gamma_\rho,\gamma_\tau]$.
Then, $\big< \frac{\delta O} {\delta \bar \xi(x)}|_{\xi = 0} \big> $, 
$\big< O \, \frac{\delta S_{GF}}{\delta \bar \xi(x)}|_{\xi = 0} \big> $ and 
$\big< O \, \frac{\delta S_{FP}}{\delta \bar \xi(x)}|_{\xi = 0} \big> $ are the contact terms,
gauge fixing terms and Faddeev-Popov terms, respectively (we do not report them here, see Ref.~\refcite{ward}).
Notice that in Eq.~(\ref{renorm}) these terms are also renormalized. 
This is because their one-loop corrections are not just multiples of the corresponding tree-level values.
Finally, $\sum_j Z_{B_j} \big< O B_j \big> $ represent the mixing not only with 
non-gauge invariant operators (in the case the operator insertion $O$ is non-gauge invariant),
but also extra mixing with gauge invariant operators which do not vanish in the off-shell regime
but vanish in the on-shell one. In principle, one require a complete list of them, or, as in our case,
a sub-list of operators whose contributions are different from zero 
to the renormalization constant we are interested on. This point will become more clear in the following.

In the supersymmetric limit, the renormalized gluino mass is zero, so the third term in the first line 
of Eq.~(\ref{renorm}) vanish and we leave with a simple expression. From now on, when 
we refer to Eq.~(\ref{renorm}) we will assume the supersymmetric limit.

\section{Renormalization constants for the supercurrent}
We are now considering each matrix element in Eq.~(\ref{renorm}) with $O$ a non-gauge invariant operator
given by 
$O := A_\nu^b(y)\, \bar \lambda^a(z) $.
Each matrix element in Eq.~(\ref{renorm}) is proportional to each element of the $Gamma$-matrix base 
$\Gamma = \left\{ 1, \gamma_5, \gamma_\alpha, \gamma_5 \gamma_\alpha, \sigma_{\alpha \rho} \right\} $.
To determine $Z_T$ one needs the projections over $\gamma_\alpha$ and $\gamma_\alpha \gamma_5$ 
(the other ones are trivial).

In Fourier transformation (FT), we choose $p$ as the outcoming momentum for the gluon field 
$A_\mu$ and $q$ the incoming momentum for the fermion field $\lambda$. Thus,
each matrix element can be written as 
$ \big< A_\nu^b(y) \, \bar \lambda^a(z) \, C(x) \big> \stackrel{FT}{\Longrightarrow}
D_F(q) \cdot (C(p,q))_{amp} \cdot D_B(p) \cdot \delta_{ab} $,
where $(C(p,q))_{amp}$ can be, i.e., $\nabla_\mu S_\mu$, $\nabla_\mu T_\mu$, etc.,  
with the external propagators amputated,
$D_F(q)$ and $D_B(p)$ are the full fermion and gluon propagators, 
respectively, while $\delta_{ab}$ is the color structure, similar to all diagrams.
The non-trivial part of the calculation is the determination of $(C(p,q))_{amp}$ for each matrix 
element in Eq.~(\ref{renorm}). To calculate $Z_T$ one should pick up
from each matrix element of Eq.~(\ref{renorm}) those terms which contains the same Lorentz structure 
as $S_\mu$ and $T_\mu$, to tree-level. Those operators which do not contain the same 
tree-level Lorentz structure as $S_\mu$ and $T_\mu$ do not enter in the determination of $Z_T$.

The renormalization constants as well as the operators, can be written as a power of $g_0$\cite{ward}, 
\beeqa
&& Z_{operator} = Z_{operator}^{(0)} + g_0^2 Z_{operator}^{(2)} + \cdots \, , \nonumber \\
&& \big< Operator \big> = \big< Operator \big>^{(0)} + g_0^2 \big< Operator \big>^{(2)} + \cdots \, ,
\label{operators}
\eneqa
where $\big< Operator \big>^{(2)}$, is the 1-loop correction
while $\big< Operator \big>^{(0)}$, is the tree-level value. It is easy to see that,
for $p = q $, a condition which would greatly simplify the calculation because implies that the operator insertion
is at zero momentum, $\big(\nabla_\mu S_\mu(x) \big)^{(0)}_{amp} = \big(\nabla_\mu T_\mu(x) \big)^{(0)}_{amp} = 0$.
That means that the tree-level of $\nabla_\mu S_\mu$ and $\nabla_\mu T_\mu$ can not be distinguished 
at zero momentum transfer.
To get $Z_T$ we need to distinguish the tree-level values of $S_\mu$ and $T_\mu$ and
for that reason we require general external momenta, $p$ and $q$.

Substituting Eq.~(\ref{operators}) into Eq.~(\ref{renorm}), up to order $g_0^2$, and
using the projections over $\gamma_\alpha$ and $\gamma_\alpha \gamma_5$ we obtain, respectively 
(where the explicit expression for the tree-level operators in Eq.~(\ref{renorm}) are used\cite{ward}), 
\beeqa
&& \frac{1}{4} \mbox{tr} \big( \gamma_\alpha \big(\nabla_\mu S_\mu \big)^{(2)}_{amp} \big) + 
Z_S^{(2)} 2 i (p_\alpha p_\nu- p_\alpha q_\nu - p^2 \delta_{\alpha \nu} + p \cdot q \delta_{\alpha \nu}) + \nonumber \\
&& Z_T^{(2)} i (p_\alpha p_\nu - p_\alpha q_\nu - p^2 \delta_{\alpha \nu} + p \cdot q \delta_{\alpha \nu}) + 
\frac{1}{4} \mbox{tr} \big(\gamma_\alpha \big(\frac{\delta O}{\delta \bar \xi(x)}|_{\xi = 0} \big)_{amp}^{(2)} \big) + \nonumber \\
&& Z_{CT}^{(2)} 2 i (p_\alpha q_\nu - p \cdot q \delta_{\alpha \nu} + p^2 \delta_{\alpha \nu} ) - 
Z_{GF}^{(2)} 2 i p_\alpha p_\nu - 
\frac{1}{4} \mbox{tr} \big( \gamma_\alpha \big(\frac{\delta S_{GF}}{\delta \bar \xi(x)}|_{\xi = 0} \big)_{amp}^{(2)} \big) 
\nonumber \\ 
&& + \frac{1}{4} Z_{B_j}^{(2)} \mbox{tr} \big< \gamma_\alpha O B_j \big>^{(0)} = 0 
\label{gammamu}
\eneqa
and 
\beeqa
&& \frac{1}{4} \mbox{tr} \big( \gamma_\alpha \gamma_5 \big( \nabla_\mu S_\mu \big)_{amp}^{(2)} \big) + 
Z_S^{(2)} 2 i p_\rho q_\sigma \varepsilon_{\nu \alpha \rho \sigma} - 
Z_{CT}^{(2)} 2 i p_\rho q_\sigma \varepsilon_{\nu \alpha \rho \sigma} + \nonumber \\
&& \frac{1}{4} \mbox{tr} \big( \gamma_\alpha \gamma_5 \big(\frac{ \delta O}{\delta \bar \xi(x)}|_{\xi = 0} \big)_{amp}^{(2)} \big) 
- \frac{1}{4} \mbox{tr} \big( \gamma_\alpha \gamma_5 \big( \frac{\delta S_{GF}}{\delta \bar \xi(x)}|_{\xi = 0} \big)_{amp}^{(2)} \big) 
+ \nonumber \\
&& \frac{1}{4} Z_{B_j}^{(2)} \mbox{tr} \big< \gamma_\alpha \gamma_5 O B_j \big>^{(0)} = 0 \, .
\label{gammamu5}
\eneqa
More explicitly, the non-trivial part is the computation of the one-loop correction of the projections over 
$\Gamma_r = \left\{ \gamma_\alpha ,\gamma_\alpha \gamma_5 \right\}$ (off-shell regime), for
$ \mbox{tr} \big( \Gamma_r \big(\nabla_\mu S_\mu \big)^{(2)}_{amp} \big)$ (12 diagrams), 
$  \mbox{tr} \big( \Gamma_r \big(\frac{\delta S_{GF}}{\delta \bar \xi(x)}|_{\xi = 0} \big)_{amp}^{(2)} 
\big) $ (12 diagrams) and 
$ \mbox{tr} \big(\Gamma_r \big(\frac{\delta O}{\delta \bar \xi(x)}|_{\xi = 0} \big)_{amp}^{(2)} \big)$ 
(4 diagrams), for a total of 28 Feynman diagrams for the determination of $Z_T^{(2)}$ (see Ref.~\refcite{ward} for details).

The determination of $Z_T^{(2)}$ require the knowledge of which operators in 
$ Z^{(2)}_{B_j} \mbox{tr} \big< \Gamma_r O B_j \big>^{(0)} $ should be included. 
Our claim\cite{ward} is that, to get $Z_T^{(2)}$ one can substitute without any ambiguity,
$\frac{1}{4} Z_{B_i}^{(2)} \mbox{tr} \big< \gamma_\alpha O B_i \big>^{(0)} \to
Z_{B_0}^{(2)} i (p_\alpha p_\nu - p^2 \delta_{\alpha \nu}) + Z_{B_1}^{(2)} i p_\nu q_\alpha +
Z_{B_2}^{(2)} i q_\nu q_\alpha + Z_{B_3}^{(2)} i q^2 \delta_{\alpha \nu} $, and 
$\frac{1}{4} Z_{B_i}^{(2)} \mbox{Tr} \big< \gamma_\alpha \gamma_5 O B_j \big>^{(0)} \to 0$.
From the explicit computation of the matrix elements of the WTi we get the following expressions,
\beeqa
\mbox{tr} \big< \gamma_\alpha \Delta \big>^{(2)} &\stackrel{FT}{\Longrightarrow} & A_1 q^2 \hat p_\alpha \hat p_\nu +
A_2 q^2 \hat p_\alpha \hat q_\nu + 
(A_3 + M_3) q^2  \delta_{\alpha \nu} + M_1 q^2 \hat p_\nu \hat q_\alpha + \nonumber \\
&&  M_2 q^2 \hat q_\alpha \hat q_\nu +  P_1 q^2 \hat p_\nu^2 \delta_{\nu \alpha} +
P_2 q^2 \hat q_\nu^2 \delta_{\nu \alpha} + \cdots
\label{eq1}
\eneqa
and
\beeq
\mbox{tr} \big< \gamma_\alpha \gamma_5 \Delta \big>^{(2)} \stackrel{FT}{\Longrightarrow}
A_4 q^2 \hat p_\rho \hat q_\sigma \varepsilon_{\nu \alpha \rho \sigma} \, ,
\label{eq2}
\eneq
where $\Delta \equiv O \nabla_\mu S_\mu(x) + \frac{ \delta O}{\delta \bar \xi(x)}|_{\xi = 0} -
O \, \frac{\delta S_{GF}}{\delta \bar \xi(x)}|_{\xi = 0} $ and the following semplification, 
$p^2=q^2$ and $ p \cdot q =0 $, has been used.
Notice that Eq.~(\ref{eq1}), even in the continuum limit, contains non-Lorentz covariant terms. 

From the matching of Eqs.~(\ref{gammamu},\ref{gammamu5}) with Eqs.~(\ref{eq1},\ref{eq2}), 
the following conditions can be derived,
\beeqa
&& \hspace{-0.4cm} A_1 = -2 i Z_S^{(2)} - i Z_T^{(2)} + 2 i Z_{GF}^{(2)} -i Z_{B_0}^{(2)} \, , \nonumber \\
&& \hspace{-0.4cm} A_3 + M_3 = 2 i Z_S^{(2)} + i Z_T^{(2)} - 2 i Z_{CT}^{(2)} + i Z_{B_0}^{(2)} -i Z_{B_3}^{(2)} \, ,\nonumber
\\
&& \hspace{-0.4cm} M_1 = -i Z_{B_1}^{(2)} \, , \hspace{0.2cm} M_2 = -i Z_{B_2}^{(2)} \, , \hspace{0.2cm} \cdots
\label{expr0}
\eneqa
and
\beeq
A_2 = 2 i Z_S^{(2)} + i Z_T^{(2)} - 2 i Z_{CT}^{(2)} \, , \hspace{0.2cm} A_4 = -2 i Z_S^{(2)} + 2 i Z_{CT}^{(2)} \, .
\label{expr1}
\eneq
The last two conditions can be explicitly solved for $Z_T^{(2)}$,
\beeq
Z_T^{(2)} = -i A_2 - i A_4 \, .
\label{expr2}
\eneq
We see from Eqs.~(\ref{expr1},\ref{expr2}) that $Z_T^{(2)}$ does not need to know the $Z_{B_j}$ which enter explicitly in
Eq.~(\ref{expr0}). That is why we did not include the operators which produce non-Lorentz covariant terms 
in Eq.~(\ref{expr0}). Their contribution vanish to the determination of $Z_T^{(2)}$.

Our perturbative result (using a Vegas Monte Carlo routine) is $Z_T^{(2)} = Z_T^{(2)}|_{1-loop} = 0.664$. 
To compare it with the numerical one\cite{farchioni}, $ Z_T^{NUM} \equiv Z_T/Z_S = 0.185(7)$,
one has to divide it by 2, $Z_T^{PT} = \frac{1}{2} Z_T^{(2)}|_{1-loop} = 0.332$.

\section{Conclusions}
A general procedure to calculate the renormalization constant $Z_T$ has been presented.
All the contribution to the supersymmetric WTi has been calculated to one-loop order in lattice perturbation theory.
Comparison with numerical results\cite{farchioni} gives good agreement. 
An important point here is that, even in the continuum limit, Lorentz breaking terms appears in Eq.~(\ref{eq1}).
The nice point is that, once the $Z_T^{(2)}$ has been determined, it is possible to impose the
on-shell condition on the gluino mass. These Lorentz breaking terms cancel out from Eq.~(\ref{eq1})
and the continuum supersymmetric WTi is recovered. The determination of $Z_S$ is under way. 

\section*{Acknowledgments}
A.~F. would like to thank H. Suganuma for the opportunity to talk at this stimulating conference.
It is also a pleasure to thank M.~Bonini, R.~De~Pietri, S.~Furui, K.~Konishi, M.~Shifman, I.-O.~Stamatescu 
for useful discussions.

\end{document}